\documentclass[aps,pre,twocolumn,showpacs,superscriptaddress,groupedaddress]{revtex4}
\usepackage{amsfonts}
\usepackage{amsmath}
\usepackage{amssymb}
\usepackage{version}
\usepackage{graphicx}
\usepackage{color} 
\includeversion{New_connection}
\excludeversion{Old_connection}

\begin{document}

\title{Coarse grained molecular simulations of membrane adhesion domains}

\author{Nadiv Dharan}
\affiliation{Department of Biomedical Engineering, Ben Gurion University,
Be'er Sheva 84105, Israel}
%\email{dharan@post.bgu.ac.il}
\author{Oded Farago}
\affiliation{Department of Biomedical Engineering, Ben Gurion University,
Be'er Sheva 84105, Israel}
\affiliation{Ilse Katz Institute for Nanoscale Science and Technology, Ben Gurion University,
Be'er Sheva 84105, Israel}
\begin{abstract}
  
We use a coarse grained molecular model of supported lipid bilayers to
study the formation of adhesion domains. We find that this process is
a first order phase transition, triggered by a combination of pairwise
short range attractive interactions between the adhesion bonds and
many-body Casimir-like interactions, mediated by the membrane thermal
undulations. The simulation results display an excellent agreement
with the recently proposed Weil-Farago 2D lattice model, in which the
occupied and empty sites represent, respectively, the adhesion bonds
and unbound segments of the membrane. A second phase transition, into
a hexatic phase, is observed when the attraction between the adhesion
bonds is further strengthened.

\end{abstract}


\maketitle

\vspace{0.45cm}

%\newpage

%\section{Introduction}
%\label{sec:intro}

Lipid membranes define the boundaries of living cells and function as
physical barriers that prevent unwanted uptake (leakage) of different
ions and molecules into (out of) the cell \cite{Alberts}. The ability
of membranes to adhere to different elements, such as the
extracellular matrix (ECM), the cytoskeleton and other membranes, is
controlled by adhesion molecules and is crucial for many biological
processes \cite{Katz}. Membrane adhesion bonds may aggregate into
large adhesion domains to provide stronger anchoring of the cell to
the ECM and other neighboring cells
\cite{Wang,Kupfer,Lenne}. Generally speaking, clustering of membrane
adhesion bonds is facilitated by several factors, such as
electrostatic and van der Walls interactions \cite{Israel}, effective
forces arising from the action of the cytoskeleton \cite{Geiger}, and
membrane mediated interactions \cite{Pincus}. In this work we put our
focus on the latter type of interactions whose origin is the entropy
associated with the thermal undulations of the membrane
\cite{Farago_book}, and which can be understood heuristically as
follows: Consider, for instance, a membrane bound to a solid surface
by several adhesion bonds. When comparing to a free (unbound)
membrane, the bound one undergoes smaller height fluctuations, thus
loosing some entropy. However, the aggregation of adhesion bonds into
a single adhesion cluster allows the unbound segments of the membrane
to fluctuate more freely, which drives the membrane to a lower free
energy state. The membrane fluctuations, thus, induce an effective
attractive interaction between the adhesion bonds.

During recent years, considerable effort has been directed toward
understanding the biophysical principles that govern the clustering
process of adhesion bonds. Traditionally, a lattice model is used, in
which the membrane is discretized into patches, which may or may not
contain adhesion molecules that bind (via receptor-ligand bonds) the
membrane to an underlying surface. Lipowsky and Weikl
\cite{Lipowsky,Weikl} proposed a model in which the system Hamiltonian
involves three terms: i) Helfrich curvature elastic energy, ii) the
energy of the specific ligand-receptor bonds, and iii) a generic
interaction term between the membrane and the surface. A closely
related model was introduced more recently by Speck and Vink
\cite{Speck}, with an additional feature of tethering the membrane at
several points (distinct from the adhesion sites) to the
cytoskeleton. Both models predict a domain formation through a
cooperative binding process, i.e., a process where the binding of a
receptor-ligand pair facilitates conditions for the formation of other
bonds in its vicinity.

The aforementioned models constitute discrete versions of Helfrich
continuum surface model of lipid bilayers. Thus, each lattice site is
characterized by two variables $s_i$ and $h_i$. The former parameter
characterizes the distribution of adhesion bonds, where $s_i=1$
corresponds to a membrane segment that is connected to the surface and
$s_i=0$ to a segment which is free to fluctuate. The latter parameter,
$h_i$, represents the local height of the membrane.  Analyzing the
aggregation behavior of the adhesion bonds by means of computer
simulations requires sampling over different distributions of lattice
sites, as well as over different height conformations. This may become
a computationally expensive task in simulations of large systems. It
is, therefore, desirable to develop a model that integrates out the
degrees of freedom associated with the height fluctuations and,
instead, assigns a potential of mean force between the lattice
adhesion sites. Apart from computational simplicity, another advantage
of this approach is that it offers direct comparison with the
well-investigated two-dimensional (2D) lattice-gas model and, thus,
highlights the role played by the membrane-mediated interactions in
the aggregation process. Such a lattice model has been recently
proposed by Weil and Farago (WF) \cite{Weil-Farago}. (We note that an
opposite approach is taken in refs.~\cite{Lipowsky,Weikl}, where the
positional degrees of freedom $s_i$ are integrated out by using the
mean field solution of the 2D lattice-gas model. This yields an
effective membrane-surface interaction energy term in the Helfrich
Hamiltonian that depends on the local $h_i$.) The WF model combines
two attractive energy terms:
\begin{equation}
{\cal H}=-\epsilon\sum_{i,j}s_is_j+\sum_{i}V_i(1-s_i).
\label{eq:WF_energy}
\end{equation}
The first term constitutes the conventional lattice-gas model, where
the sum runs over all pairs of nearest neighbor sites. The energy
$\epsilon>0$ gained for each pair of occupied sites accounts for all
the interactions between the adhesion bonds other than the
membrane-mediated potential of mean force. The latter potential is
represented by the second term in Eq.~(\ref{eq:WF_energy}) which,
quite unusually, involves summation over the {\em empty}\/ sites
only. The energy of each empty site measures the amount of free energy
lost due to the suppression of the thermal height fluctuations of the
corresponding membrane segment. Weil and Farago conjectured that this
free energy penalty depends on the distance of the segment from the
nearest adhesion bond $d_i^{\rm min}$, i.e., the distance to the
nearest occupied site, and is given by
\begin{equation}
V_i=\frac{k_{\rm B}T}{\pi}\left(\frac{l}{d_i^{\rm min}}\right)^2,
\label{eq:Vr}
\end{equation}
where $l$ is the lattice constant (which should be of the order of a
few nanometers - comparable to the thickness of the
membrane). Remarkably, the expression for the free energy $V_i$
(\ref{eq:Vr}) is independent of the bending rigidity of the membrane
$\kappa$. Notice that, in general, $d_i^{\rm min}$ depends on the
distribution of {\em all}\/ the occupied sites and, therefore, the
second term in Eq.~(\ref{eq:WF_energy}) represents a multi-body
potential of mean force between the adhesion bonds. This potential is
attractive because most of the entropy is lost at the interfacial
regions between occupied and empty sites where $d_i^{\rm min}$ is
small.  When only two sites are occupied, the potential between them
has a logarithmic dependence on their separation $r$:
\begin{equation}
U(r)=2k_{\rm B}T \ln \left(\frac{r}{l}\right ).
\label{eq:PMF2}
\end{equation}
The last result has been obtained independently through scaling
arguments and has been verified by computer simulations of
coarse-grained bilayer membranes \cite{Farago}. 

Monte Carlo (MC) simulations of the WF model reveal that the system
condenses for $\epsilon>\epsilon_c>0$. The transition value,
$\epsilon_c$, is smaller than the corresponding value of the standard
lattice-gas model at the same density by typically a factor of
2-3. Noticeably, $\epsilon_c$ is smaller than thermal energy $k_{\rm
B}T$ in the WF model, and larger than $k_{\rm B}T$ in the standard
lattice-gas model. In agreement with previous lattice models that
include the membrane explicitly (and not via a potential of mean
force) \cite{Lipowsky,Weikl,Speck}, the adhesion sites do not form
large clusters when $\epsilon=0$, which implies that the
membrane-mediated interactions alone are not sufficient to allow the
formation of large adhesion domains, but they greatly reduce the
strength of the residual interactions required to facilitate cluster
formation. Following this study, Noguchi suggested that the strength
of the membrane-mediated interactions can be enhanced by pinning more
than one membrane to the surface \cite{Noguchi}. He demonstrated this
by simulating monolayers of particles that are pinned to each other by
``gap junctions''. In simulations of $N_{\rm lay}=2$, the gap
junctions remain dispersed. This result agrees with the prediction of
the WF model for $\epsilon=0$ since the problem of two surfaces with
bending rigidity $\kappa$ is equivalent to a single membrane with
$\kappa/2$ connected to an infinitely rigid surface. However, when the
number of monolayers is $N_{\rm lay}>2$, the gap junctions exhibit a
different behavior and condense into a large stable domain. This
behavior can be attributed to the fact that the entropy loss caused by
the gap junctions is proportional to the total rate of collisions
between the layers in the stack \cite{Farago}, which grows
proportionally to the number of {\em pairs}\/ of colliding surfaces,
i.e., to $(N_{\rm lay}-1)$. Motivated by the results of the molecular
simulations, Noguchi also simulated the WF lattice model, with a free
energy term which is simply $\left (N_{\rm lay}-1\right )$ times
larger than $V_i$ given by Eq.~(\ref{eq:Vr}). It was found that the WF
model yields results in very good agreement with the molecular
simulations.

In this paper we provide yet another evidence for the ability of the
WF model to accurately capture the aggregation behavior of adhesion
bonds in supported membranes. To this end, we use the model proposed
by Cooke and Deserno (CD), in which lipids are modeled as trimmers
consisting of one hydrophilic (head) and two hydrophobic (tail) beads
\cite{Deserno}. This model is less coarse-grained than the one used by
Noguchi and, thus, gives a better representation of lipid membranes
which are simulated as bilayers rather than monolayers. A flat plate,
which cannot be intersected by the lipids, was placed underneath the
lower monolayer at $z=0$, and the attachment of the membrane to the
surface was established by restricting $N_b$ head beads from the lower
monolayer to $z=0$ and allowing them to move only in-plane.  We
conducted MC simulations with periodic boundary conditions of a
bilayer comprising of $2N=2000$ lipids at different densities of
adhesive lipids, $\phi=N_b/N$.  A slight change in the CD model was
made where, for pairs of adhesive head beads, the pair potential was
switched from head-head to tail-tail. While the former pair potential
is purely repulsive, the latter also includes a cosine potential well
whose depth can be tuned (see Eq.~(4) in ref.~\cite{Deserno}). This
attractive part of the pair potential plays the same role played by
the standard lattice-gas term in Eq.~(\ref{eq:WF_energy}), with
$\epsilon$ denoting the interaction energy between nearest neighbor
occupied sites.  By setting the depth of the potential well in the
molecular model to $\epsilon$, and by simulating the WF lattice model
with same value of $\epsilon$, one can directly compare the two models
to each other. This allows us to test the accuracy of the WF model for
$\epsilon>0$ - an aspect of the model which has not been probed in
ref.~\cite{Noguchi}.

The simulations of the CD model (to be henceforth referred to as the
``molecular simulations''), which were conducted at zero surface
tension, consist of several types of MC moves, including translation
of beads, rotation of lipids, and changes in the cross-sectional
projected area of the membrane.  To achieve equilibration within a
reasonable computing time, two additional move types were also
performed. The first move type resolves the problem arising from the
slow changes in the amplitudes of the large wavelength bending
modes~\cite{Farago_mode}. It involves a collective change in the
heights of all the lipids, allowing acceleration and rapid relaxation
of these modes. The other process limiting the approach to equilibrium
is the slow diffusion of the lipids, especially those pinned to the
surface which serve as the adhesion bonds. In order to speed up the
aggregation of adhesion domains, one needs to allow the adhesion bonds
to ``jump'' across the membrane. This is accomplished by the second
move type, in which two lipids simultaneously experience opposite
vertical translations: the free lipid whose head resides closest to
the surface is brought down and attached to the surface, while a
randomly chosen pinned lipid is lifted and released~\cite{Farago}.

\begin{figure}[t]
\begin{center}
\scalebox{0.275}{\centering\includegraphics{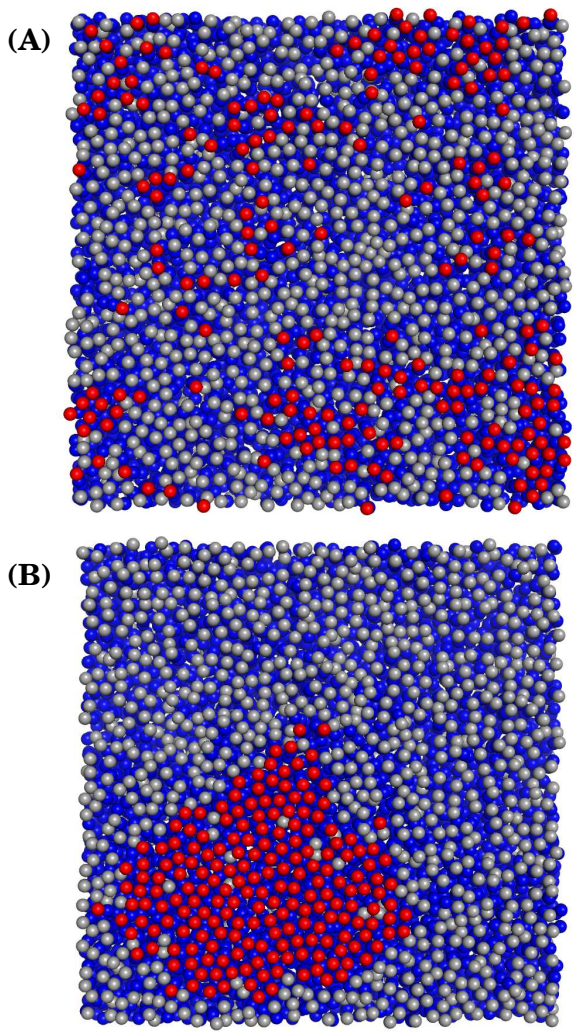}}
\end{center}
\vspace{-0.5cm}
\caption{Bottom view of a membrane with concentration of adhesion
bonds $\phi=0.2$ for (A) $\epsilon=0.4$ and (B) $\epsilon=1.2$. The
head and tail beads of the lipids are colored in grey and blue,
respectively, while the adhesive beads are colored in red}.
\label{fig:fig1}
\end{figure}

\begin{figure}[b]
\begin{center}
\scalebox{0.44}{\centering\includegraphics{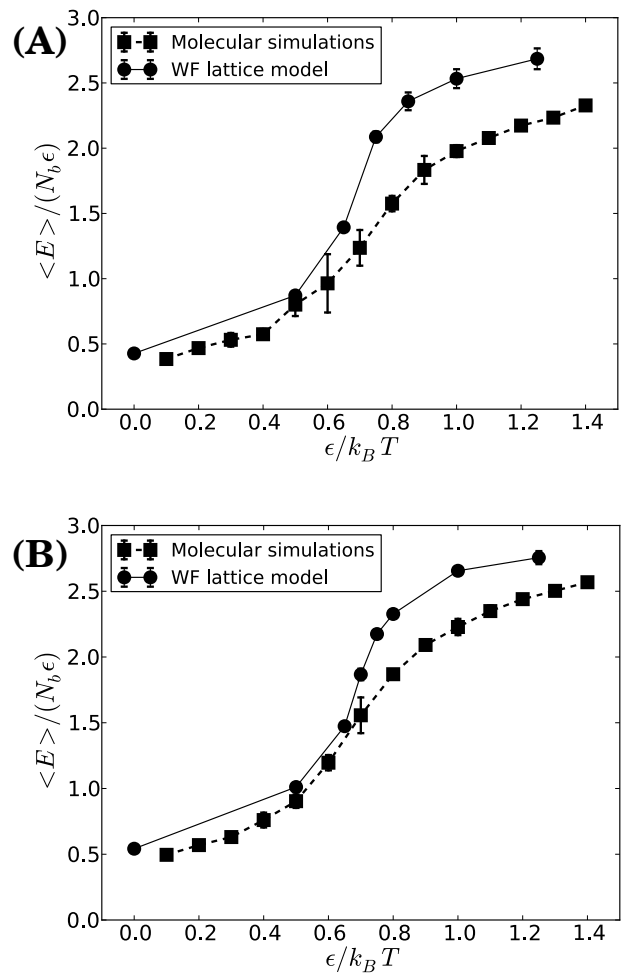}}
\end{center}
\vspace{-0.5cm}
\caption{The average energy of direct interactions between the
 adhesion bonds (normalized per adhesion bond) as function of the pair
 interaction energy $\epsilon$, for (A) $\phi=0.05$ and (B)
 $\phi=0.1$. Solid squares and circles denote the results of the
 molecular simulations and of the Weil-Farago 2D lattice simulations,
 respectively. The solid and dashed lines are guides to the eye.}
\label{fig:fig2}
\end{figure}

We simulated membranes with different concentrations $\phi$ of
adhesion bonds, and for different values of $\epsilon$ (measured in
units of the thermal energy $k_{\rm B}T$). Snapshots of equilibrium
configurations corresponding to $\epsilon=0.4$ and $\epsilon=1.2$ are
shown, respectively, in Figs.~\ref{fig:fig1}(A) and
Fig.~\ref{fig:fig1}(B). The concentration in both cases is
$\phi=0.2$. The distinction between the two configurations is clear:
In (A) the adhesion bonds are scattered across the membrane in
relatively small clusters, while in (B) they are assembled into one
big aggregate. The transition between the gas and the condense phases
of adhesion bonds displayed, respectively, in Figs.~\ref{fig:fig1}(A)
and (B) occurs at intermediate values of $\epsilon$. This is
demonstrated in Fig.~\ref{fig:fig2}, where we plot the average energy
due to pair interactions between the adhesion bonds (normalized per
bond), $\langle E\rangle/N_b$, as a function of $\epsilon$, the
(maximum) strength of the pair interaction, for $\phi=0.05$ (A) and
$\phi=0.1$ (B). The simulation results, which are plotted in solid
squares (with the dashed line serving as a guide to the eye), suggest
that the transition between the phases is of first order. The energy
steeply increases around $\epsilon_c\approx0.7$ from a low value
reflecting the dispersed distribution of adhesion bonds in the gas
phase where the number of pair interactions is small, to a high value
characterizing a big cluster where the bonds are closely packed and
experience a large number of pair interactions. Also plotted in
Fig.~\ref{fig:fig2} are the results of lattice simulations of the WF
model for identical values of $\phi$ and for various values of
$\epsilon$ (solid circles with solid line serving as a guide to the
eye). The agreement between the molecular simulations and the lattice
simulations of the WF model is very good. The lattice model predicts a
very similar value of $\epsilon_c\approx0.7$ (for both simulated
concentrations), and gives very similar values of $\left \langle
E\right \rangle/N_b$ in the gas phase ($\epsilon<\epsilon_c$).

\begin{widetext}

\begin{figure}[t]
\begin{center}
\scalebox{0.35}{\centering\includegraphics{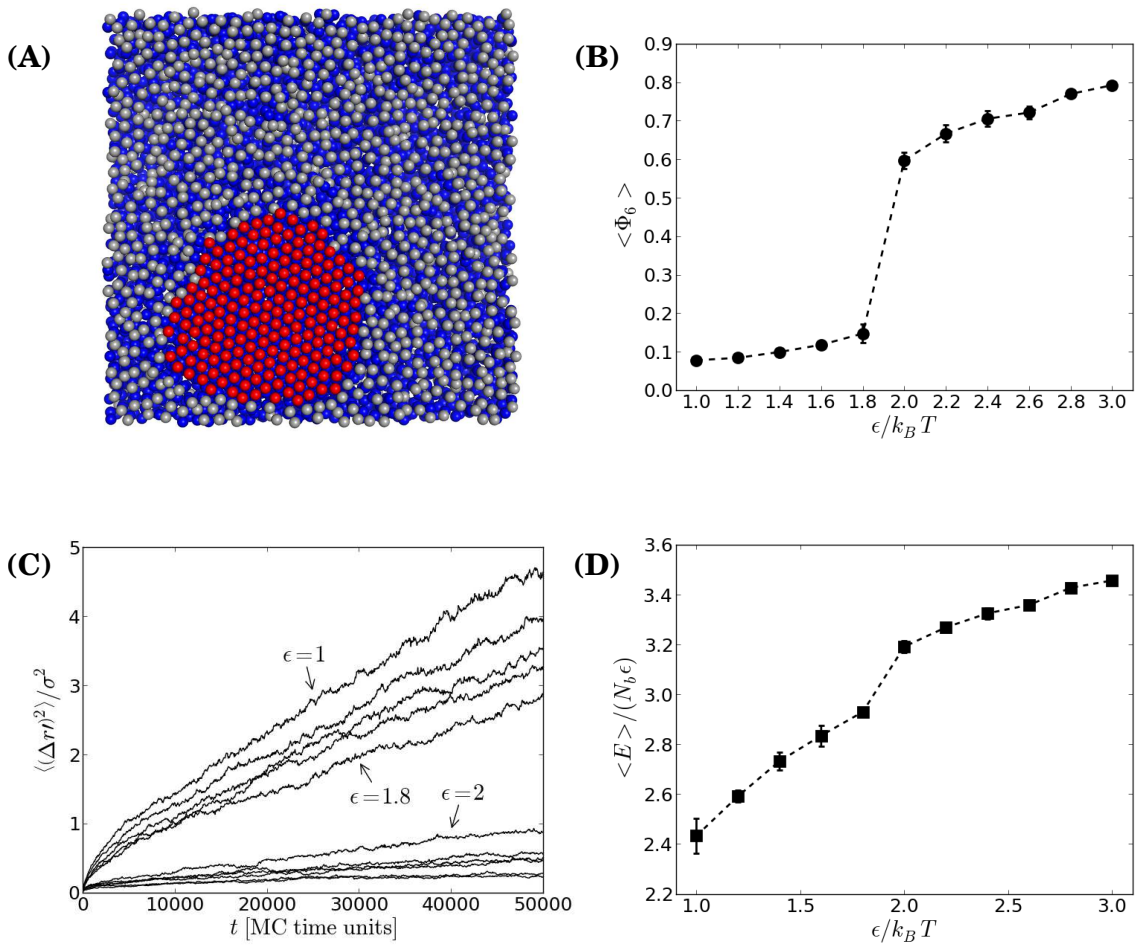}}
\end{center}
\vspace{-0.5cm}
\caption{The molecular simulation results for a membrane with
$\phi=0.2$ for $\epsilon>\epsilon_c$.(A) Snapshot of an equilibrium
configuration with $\epsilon=3.4$, depicting an adhesion domain
organized in the hexatic phase. Color coding as in Fig.~{\protect
\ref{fig:fig1}}. (B) The mean bond orientational order parameter
$\langle\Phi_6\rangle$ as a function of the pair interaction energy
$\epsilon$. The transition into the hexatic phase occurs around
$\epsilon_h\approx1.9$ where a sudden increase in
$\langle\Phi_6\rangle$ is observed. (C) The mean square displacement
of the adhesion bonds vs. the simulations time for different values of
$\epsilon$. The slope of each curve is a measure for the self
diffusion coefficient of the adhesion bonds within the cluster
$D$. The results for $\epsilon=1,1.8,2$ are marked by arrows. (D) The
average interactions energy per adhesion bond as a function of
$\epsilon$.}
\label{fig:fig3}
\end{figure}

\end{widetext}

A slight discrepancy between the molecular and lattice simulation is
observed in the condensed phase for $\epsilon>\epsilon_c$, where the
WF model appears to give higher values of the mean interaction energy
$\left \langle E\right \rangle/N_b$. This deviation between the
results of the lattice and continuum molecular models is anticipated
considering the nature of the models. In the former, the sites are
organized on a perfect triangular lattice, and the energy assigned to
every pair of nearest neighbor occupied sites is {\em exactly}\/
$\epsilon$. In the latter, on the other hand, the bonds within each
cluster do not necessarily have a long range positional order [see,
e.g., the snapshot in Fig.~\ref{fig:fig1}(B)], and $\epsilon$ denotes
the {\em depth}\/ of the interaction well. The actual strength of the
interaction is expected to be lower than $\epsilon$ in the continuum
molecular model, which explains why it gives lower values of $\left
\langle E\right \rangle/N_b$ than in the lattice simulations.

At even higher values of $\epsilon$, the close agreement between the
lattice and the molecular simulations is regained. This occurs due to
another phase transition that the clusters undergo, from disordered
liquid-like structures into more ordered organizations such as the one
displayed in Fig.~\ref{fig:fig3}(A) for $\phi=0.2$ and $\epsilon=3.4$.
This phase transition can be understood within the framework of the
KTHNY theory, which proposes the formation of a two dimensional
hexatic phase with a quasi-long range hexagonal (orientational) order
\cite{KTHNY}. This transition is characterized by the bond
orientational order parameter
 \begin{equation}
\psi_{6j}=\frac{1}{N_j}\sum_{k=1}^{N_j}e^{i6\theta_{kj}},
\label{eq:loc_phi6}
\end{equation}
where the sum runs over the nearest neighbor bonds $k$ to a given bond
$j$ (whose identity is determined by Voronoi tessellation), and
$\theta_{kj}$ is the angle between the line connecting the pair of
bonds $j$ and $k$ and some fixed axis. Averaging over all the bonds
within the cluster yields the global orientational order parameter
\begin{equation}
\Phi_6=\left |\frac{1}{N_b}\sum_{j=1}^{N_b}\psi_{6j}\right |.
\label{eq:glob_phi6}
\end{equation}
Another quantity undergoing rapid variations at the transition is the
self-diffusion coefficient of the bonds (relative to the diffusion of
their center of mass), defined by
\begin{multline}
D=\lim_{t\rightarrow \infty }\frac{1}{4N_bt}\sum_{i=1}^{N_b}
\Big\langle\left[\left(\vec{r}_i(t)-\vec{r}_{\rm cm}(t)\right)-\right. 
 \\ \left. \left. \left(\vec{r}_i(t=0)-\vec{r}_{\rm
cm}(t=0)\right)\right]^2\right\rangle\equiv\lim_{t\rightarrow \infty
}\frac{\left\langle\left( \Delta r\prime\right)^2\right\rangle}{4t} ,
\label{eq:self_difus}
\end{multline}
where $r_i(t)$ and $r_{\rm cm}(t)$ denote, respectively, the position
of adhesion bond $i$ and of the center of mass of the cluster at time
$t$ (measured in MC time units), and $\langle\cdots\rangle$ denotes
statistical average. The transition into the hexatic phase is
characterized by (i) an increase in $\Phi_6$, associated with the
emrgence of orientational order, and (ii) a sharp decrease in $D$,
reflecting a lower mobility of the bonds.  In Fig.~\ref{fig:fig3}(B),
we plot our results for $\langle\Phi_6\rangle$, as a function of
$\epsilon$ for $\phi=0.2$. In Fig.~\ref{fig:fig3}(C) the mean square
displacement of the adhesion bonds (measured in units of $\sigma^2$,
where $\sigma$ is the range of the head-head repulsive potential in
the Cooke-Deserno model \cite{Deserno}) is plotted versus the
simulation time (measured in MC time units), with the curves, from top
to bottom, corresponding to increasingly higher values of $\epsilon$.
[Each curve in Fig.~\ref{fig:fig3}(C) corresponds to a data point in
Fig.~\ref{fig:fig3}(B)].  The curves display a linear increase in
$\langle( \Delta r\prime)^2\rangle$ with $t$, and the slope of each
curve is proportional to $D$. Both Figs.~\ref{fig:fig3}(B) and (C)
indicate that the transition from disorder-liquid into an
ordered-hexatic structure occurs at around $\epsilon\approx
1.9$. Another evidence for the fluid to hexatic transition is also
observed in Fig~\ref{fig:fig3}(D), showing a ``jump'' in the average
interaction energy between $\epsilon=1,8$ and $\epsilon=2.0$. Notice
that the values of $\langle E\rangle/N_b$ in the hexatic phase is {\em
higher}\/ the 3, which is the maximum possible value in simulations of
the WF model on a triangular lattice. This feature is related to the
form of the attractive tail-tail pair potential in the molecular
simulations whose cut-off range was set to slightly less that
$2.5\sigma$. This implies that, in a closely packed cluster, each
adhesion bond may weakly interact with its next- and next-next-nearest
neighbors, which explains why $\langle E\rangle/N_b$ is larger than
$\epsilon$.

To conclude, we used coarse-grained molecular simulations to study the
aggregation of adhesion domains in supported membranes. Formation of
adhesion domains occurs due to two types of attractive interactions
existing between the adhesion bonds. These include (i) a many-body
potential of mean force induced by the thermal fluctuations of the
membrane, and (ii) short-range pair interactions of strength
$\epsilon$. Upon increasing $\epsilon$, the system goes from a ``gas''
phase where the bonds are scattered across the membrane in relatively
small clusters, into a ``condensed'' fluid phase, in which they are
assembled into large aggregates. At even higher values of $\epsilon$,
another phase transition is observed from a condensed fluid-like phase
into a more ordered hexatic phase, in which the bonds also exhibit a
considerably reduced diffusivity. Based on our computational
observations, we respectively identify these transitions as a
first-order condensation transition, and a Kosterlitz-Thouless phase
transition. To rigorously characterize the nature of the transitions,
one would need to perform finite size scaling analysis, but this goes
beyond the scope of the current study.

Our simulation results, especially those related to the condensation
transition, appear to be in excellent agreement with the recently
proposed Weil-Farago (WF) lattice model. This lands credibility to the
main idea of the WF model, which is to associate the
fluctuation-induced potential between the bonds, with free energies
assigned to the empty sites of the lattice. The empty sites represent
the fluctuating segments of the supported membrane, and the free
energy assigned to each site measures the free energy loss due to the
local restrictions imposed on the membrane thermal undulations. This
free energy penalty mainly depends on the distance, $d_{\rm min}$,
between an empty site and the closest occupied site (representing an
adhesion bond). In the present paper, we investigated tensionless
membranes with adhesion bonds directly pinned to the underlying
surface. Under these conditions, the WF model assumes that the energy
of the empty sites scales proportionally to $d_{\rm min}^{-2}$. In a
future publication we plan to extend the WF lattice model to more
general conditions. The extensions of the WF model will be tested against
molecular simulations akin to those presented here.

Acknowledgments: This work was supported by the Israel Science
Foundation through grant No.~1087/13.

% References

\end{document}